\documentclass[10pt, nolongbibliography,groupedaddress,showpacs,showkeys,amsmath,amssymb,eqsecnum,aps,prd,nofootinbib]{revtex4-2}

\usepackage{tensor, mathtools, bm} 

\usepackage{xcolor}                   
\usepackage[caption=false]{subfig}

\usepackage{lmodern}

\usepackage[]{graphicx} 

\usepackage{hyperref}
\begin{document}

\title{A Renormalizable Model of Quantized Gravitational and Matter Fields}

\author{D. G. C. McKeon}
\email{dgmckeo2@uwo.ca}
\affiliation{Department of Applied Mathematics, The University of Western Ontario, London, Ontario N6A 5B7, Canada}
\affiliation{Department of Mathematics and Computer Science, Algoma University,
Sault Ste.~Marie, Ontario P6A 2G4, Canada}

\author{F. T. Brandt}  
\email{fbrandt@usp.br}
\affiliation{Instituto de F\'{\i}sica, Universidade de S\~ao Paulo, S\~ao Paulo, SP 05508-090, Brazil}

\author{J. Frenkel}
\email{jfrenkel@if.usp.br}
\affiliation{Instituto de F\'{\i}sica, Universidade de S\~ao Paulo, S\~ao Paulo, SP 05508-090, Brazil}

 \author{S. Martins-Filho}   
\email{sergiomartinsfilho@usp.br}
\affiliation{Instituto de F\'{\i}sica, Universidade de S\~ao Paulo, S\~ao Paulo, SP 05508-090, Brazil}

\date{\today}

\begin{abstract}
A Lagrange multiplier field can be used to restrict radiative corrections to the Einstein-Hilbert action to one-loop order. This result is employed to show that it is possible to couple a scalar field to the metric (graviton) field in such a way that the model is both renormalizable and unitary. The usual Einstein equations of motion for the gravitational field are recovered, perturbatively, in the classical limit. By evaluating the generating functional of proper Green's functions in closed form, one obtains a novel analytic contribution to the effective action.
\end{abstract}
\keywords{quantum gravity, Lagrange multipliers, renormalization}

\maketitle

\section{Introduction}
It has long been known that when the equations of motion are satisfied, the divergences that arise at one-loop order in the Einstein-Hilbert effective action can be absorbed by a field renormalization \cite{ref1, ref2, ref3}. However, once the metric couples to a scalar \cite{ref1, ref2}, vector \cite{ref4}, or spinor field \cite{ref5}, or one goes to two-loop order \cite{ref6, ref7}, this is no longer true.
In order to eliminate such unwanted divergences, it has been suggested that by extending the number of physical fields through supersymmetric coupling (supergravity) \cite{ref8}, extending the concept of particles to strings \cite{ref9}, adjusting the canonical structure of the theory (loop gravity) \cite{ref10}, invoking non-perturbative behavior (asymptotic safety) \cite{ref11, newref17}, or by including additional terms into the basic Einstein-Hilbert action \cite{ref12, ref13}, one might obtain a realistic theory of quantum gravity. 
Although much progress has been made along these lines, these approaches have not yet proven to be entirely satisfactory.

It has been known for some time that by using a Lagrange Multiplier (LM) field to ensure that the classical equations of motion are satisfied, radiative corrections to the classical action in the full effective action are restricted to one-loop order \cite{ref14}. This suggests a way of  quantizing the Einstein-Hilbert Action \cite{ref15, ref16, ref17}, with additional contributions from a scalar field \cite{ref18}. In this manner, the usual classical action is recovered at tree level and unwanted divergences are all removed through renormalization, without the introduction of extra fields or dimensions, or compromising unitarity.

In this paper, we provide details of how a model in which the metric couples to a scalar can have both satisfactory classical and quantum properties through use of an LM field. This is a continuation of the work in ref. \cite{ref18}.

In the next section, we use a standard integral to illustrate how an LM field can be used to eliminate radiative effects beyond one-loop order in perturbation theory. In Section III, a general discussion of how this approach can be used in field theory is presented. We note here the presence of a supplementary solution to the classical equations of motion which may possibly simulate ``dark matter'' effects. In Section IV, we show how the metric interacting with a scalar field can be treated so as to have a renormalizable and unitary theory. A brief discussion of the results is given in Section V. In Appendix A, we exemplify, in a scalar model, the unitarity of the theory by showing that the cross sections are related to the imaginary part of forward scattering amplitudes.

\section{A Simple Illustration}

Let us begin with the simple integral:
\begin{equation}
    Z = \int \frac{dg}{2\pi} \frac{d\lambda}{2\pi} \exp \left[i \left( L(g) + \lambda L'(g) \right)\right]. \label{eq1}
\end{equation}

By integrating first over \(\lambda\), then using the resulting \(\delta\)-function to integrate over \(g\), we obtain:
\begin{equation}
    Z = \frac{1}{2\pi} \sum_A \frac{\exp \left( i L(g^A) \right)}{|L''(g^A)|}, \label{eq2}
\end{equation}
where \(g^A\) is a solution to
\begin{equation}
    L'(g^A) = 0. \label{eq3}
\end{equation}

We can reinterpret \(g\) as a ``quantum field,'' \(L(g)\) as a ``classical action,'' \(\lambda\) as a LM field, \(g^A\) as a solution to the classical equation of motion, \(\exp \left( i L''(g^A) \right)\) as the sum of tree-level diagrams, and \(\left[ L''(g^A) \right]^{-1/2}\) as ``one-loop diagrams with external legs on the mass shell,'' and \(Z\) itself as the generating functional.

We now modify \(Z\) so that \cite{ref19, ref20}:
\begin{equation}
\begin{aligned}
    Z &= \int \frac{dg}{2\pi} \frac{d\lambda}{2\pi} \left( L''(g) \right)^{1/2} \exp \left( i \left[ L(g) + \lambda L'(g) \right] \right) \\
    &= \int \frac{dg}{2\pi} \left( L''(g) \right)^{1/2} \delta(L'(g)) \exp \left[ i L(g) \right]. \label{eq4}
\end{aligned}
\end{equation}

Having added the factor of \((L''(g))^{1/2}\) to the integrand of Eq.~\eqref{eq1}, one can show explicitly that the form of \(Z\) now remains invariant under a change of variables in which \(g'= g'(g)\).\footnote{See, however, Ref. \cite{ref36}.}

Sources \(j\) and \(J\) are now introduced so that:
\begin{eqnarray}\label{eq5}
    Z(j, J) &=& \exp \{i W(j, J)\}
\nonumber \\
  &=& \int \frac{dg}{2\pi} \frac{d\lambda}{2\pi} (L''(g))^{1/2} \exp \{i (L(g) + \lambda L'(g) + j g + J \lambda)\}.
\end{eqnarray}

``Background fields'' \(\bar{g}\) and \(\bar{\lambda}\) are defined so that \cite{ref21, ref22}:
\begin{equation} \label{eq6}
    \bar{g} = \frac{\partial W(j, J)}{\partial j}, \quad \bar{\lambda} = \frac{\partial W(j, J)}{\partial J},
\end{equation}
and a Legendre transform leads to
\[
    \Gamma(\bar{g}, \bar{\lambda}) = W(j, J) - j \bar{g} - J \bar{\lambda},
\]
where
\begin{equation} \label{eq7}
    \exp \{i \Gamma(\bar{g}, \bar{\lambda})\} = \int \frac{dh}{2\pi} \frac{dH}{2\pi} (L''(\bar{g} + h))^{1/2} \exp \{i (L(\bar{g} + h) + (\bar{\lambda} + H) L'(\bar{g} + h) + j h  + J H)\}.
\end{equation}
with
\begin{equation} \label{eq8}
    j = -\frac{\partial \Gamma(\bar{g}, \bar{\lambda})}{\partial \bar{g}}, \quad J = -\frac{\partial \Gamma(\bar{g}, \bar{\lambda})}{\partial \bar{\lambda}}.
\end{equation}

We now can set the background field for the LM field equal to zero (ie. $ \bar{\lambda} =0$) as physical contributions to one-particle irreducible (1PI) Green's functions (GF) are computed by expanding $ \Gamma (\bar{g} , 0 )$ in powers of $ \bar{g} $.

\section{General Field Theory Considerations}
First, let us recall some of the features of ``background field quantization'' \cite{ref1, ref21, ref22}. If we begin with the generating functional \(Z(j)\) for a field \(\chi(x)\) with a Lagrangian \(\mathcal{L}(\chi)\) and source \(j(x)\), then upon restoring Planck's constant \(\hbar\), we have
\begin{equation} \label{eq9}
    Z(j) = e^{\frac{i}{\hbar} W(j)} =\int \mathcal{D}\chi \exp \left\{ \frac{i}{\hbar} \int dx (\mathcal{L}(\chi) + j(x) \chi(x)) \right\}.
\end{equation}

A background field \(\bar{\chi}(x)\) is given by
\begin{equation} \label{eq10}
    \bar{\chi}(x) = \frac{\delta W(j)}{\delta j(x)},
\end{equation}
and through a Legendre transform
\begin{equation} \label{eq11}
    \Gamma(\bar{\chi}) = W(j) - \int dx \, j(x) \bar{\chi}(x), \quad \left(j(x) = -\frac{\delta \Gamma(\bar{\chi})}{\delta \bar{\chi}(x)}\right),
\end{equation}
we obtain
\begin{equation} \label{eq12}
    \exp \left\{ \frac{i}{\hbar} \Gamma(\bar\chi) \right\} = \int \mathcal{D}q \exp \left\{ \frac{i}{\hbar} \int dx \left( \mathcal{L}(\bar{\chi} + q) + j(x) q(x) \right) \right\}
\end{equation}
upon letting \(\chi(x) = \bar{\chi}(x) + q(x)\). It has been shown \cite{ref22} that \(W(j)\) is the generating functional for all connected Feynman diagrams and that \(\Gamma(\bar{\chi})\) is the generating functional for all connected one-particle irreducible diagrams. If we make the expansions
\begin{equation} \label{eq13}
    \Gamma(\bar{{\chi}}) = \Gamma^{(0)}(\bar{\chi}) + \hbar \Gamma^{(1)}(\bar{\chi}) + \hbar^2 \Gamma^{(2)}(\bar{\chi}) + \cdots,
\end{equation}
and
\begin{equation} \label{eq14}
    \mathcal{L}(\bar{\chi} + q) = \mathcal{L}(\bar{\chi}) + \mathcal{L}'(\bar{\chi}) q + \frac{1}{2!} \mathcal{L}''(\bar{\chi}) q^2 + \frac{1}{3!} \mathcal{L}'''(\bar{\chi}) q^3 + \cdots ,
\end{equation}
then, by Eq.~\eqref{eq12}, we find that
\begin{equation} \label{eq15}
    \exp \left( \frac{i}{\hbar} \Gamma^{(0)}(\bar{\chi}) \right) = \exp \left( \frac{i}{\hbar} \int dx \, \mathcal{L}(\bar{\chi}) \right),
\end{equation}
and
\begin{equation} \label{eq16}
    \exp \left( i \Gamma^{(1)}(\bar{\chi}) \right) = \det \left( \frac{\partial^2 \mathcal{L}(\bar{\chi})}{\partial \bar{\chi}(x) \partial \bar{\chi}(y)} \right)^{-1/2}.
\end{equation}
with the term  \(\mathcal{L}'(\bar{\chi}) q\) cancelling. The contributions to \(\Gamma^{(n)}(\bar{\chi})\) (\(n > 2\)) arise from one-particle irreducible Feynman diagrams with \(n\)-loops involving vertices derived from vertex terms in the expansion of Eq.~\eqref{eq14} containing at least three powers of \(q(x)\), and propagators arising from \((\mathcal{L}''(\bar{\chi})) q^2\).
It is important to note that the equation of motion for the background field $ \bar{\chi} $
\begin{equation} \label{eq17}
    \mathcal{L}'(\bar{\chi}) = 0,
\end{equation}
is not used to derive this expansion for \(\Gamma(\bar\chi)\).

The general development outlined in the preceding section can be directly applied to a model with a classical Lagrangian \(\mathcal{L}_\text{cl}(g_i(x))\) and action
\begin{equation} \label{eq18}
    S_\text{cl}(g_i, \lambda_i) = \int dx \left[ \mathcal{L}_\text{cl}(g_i(x)) + \lambda_i(x) \frac{\delta \mathcal{L}_\text{cl}(g_i(x))}{\delta g_i(x)} \right].
\end{equation}
Paralleling the way in which Eq.~\eqref{eq7} is obtained, we find that the generating functional \(\Gamma(\bar{g_i}(x), \bar{\lambda_i}(x))\) for one-particle irreducible diagrams is given by
\begin{equation} \label{eq19}
\begin{split}
    \exp \left\{ i \Gamma(\bar{g}_i(x), \bar{\lambda}_i(x)) \right\} ={}& 
     \int \mathcal{D}h_i(x) \mathcal{D}H_i(x) \det \left( \frac{\delta^2 \mathcal{L}_\text{cl}(\bar{g}_i(x) + h_i(x))}{\delta h_i(x) \delta h_j(y)} \right)^{1/2}
    \\ & \quad   \times \exp i\int dx \Big[ \mathcal{L}_\text{cl}(\bar{g}_i(x) + h_i(x))
 + \left(\bar{\lambda}_i(x) + H_i(x)\right) \frac{\delta \mathcal{L}_\text{cl}(\bar{g}_i(x) + h_i(x))}{\delta h_i(x)} 
    \\ & \quad + j_i(x) h_i(x) + J_i(x) H_i(x) \Big].
\end{split}
    \end{equation}
as in Eq.~\eqref{eq12}.
We now set the source functions and background for the LM field in Eq.~\eqref{eq19} equal to zero,
\begin{equation} \label{eq20}
    \bar{\lambda}_i = J_i = 0,
  \end{equation}
  since we are only interested in GF involving external fields $ \bar{g}_{i} $.

In ref.~\cite{ref20}, the situation in which the classical action 
\(\int dx \, \mathcal{L}_\text{cl}(g_i(x))\) is invariant under the gauge transformations
\begin{equation} \label{eq21}
    g_i \to g_i + R_{ij}(g) \xi_j
\end{equation}
is considered. This is accompanied by transformations of the LM field \(\lambda_i(x)\) as well as of the two fermionic and one bosonic ghost fields used to exponentiate the functional determinant 
\(\det^{1/2} \left( \frac{\delta^2 \mathcal{L}_\text{cl}}{\delta \phi_i \delta \phi_j} \right)\). It is shown in ref.~\cite{ref20} that the imposition of the gauge conditions
\begin{equation} \label{eq22}
    F_{ij} g_j = 0 = F_{ij} \lambda_j
\end{equation}
leads to a simple Faddeev-Popov (FP) determinant \cite{ref21, ref23},
\begin{equation} \label{eq23}
    \det \left( F_{ij} R_{jk}(g_i) \right),
\end{equation}
as well as a gauge-fixing contribution to the action
\begin{equation} \label{eq24}
    \int dx \, \mathcal{L}_\text{gf} = \int dx \, \left[ -\frac{1}{2\alpha} \left( F_{ij} g_j \right)^2 - \frac{1}{\alpha} \left( F_{ij} \lambda_j \right) ( F_{ik} g_k) \right].
\end{equation}

If one were to choose the gauge fixing to leave background gauge invariance unbroken, then
\begin{equation} \label{eq25}
    \bar{g}_i \to \bar{g}_i + R_{ij}(\bar{g}) \xi_j.
\end{equation}
remains a symmetry, and the gauge condition is now of the form
\begin{equation} \label{eq26}
    F_{ij}(\bar{g}) h_j = 0 = F_{ij}(\bar{g}) H_j,
\end{equation}
so that the FP determinant is given by
\[
    \det \left( F_{ij}(\bar{g}) R_{jk}(\bar{g} + h) \right).
\]

The generating functional in Eq.~\eqref{eq19} is now
\begin{equation} \label{eq27}
\begin{split}
    \exp \{ i \Gamma(\bar{g}_i, \bar{\lambda}) \} 
    &=  \int \mathcal{D}h_i \mathcal{D}H_i \, {\det}^{1/2} \left[ \frac{\delta^2 \left(\mathcal{L}_\text{cl}(\bar{g}_i + h_i)
    -\frac{1}{2\alpha}(F_{ij}(\bar{g}_i) h_j)^2 \right)}{\delta h_i(x) \delta h_j(y)} \right] 
\det \left( F_{ij}(\bar{g}_i) R_{jk}(\bar{g}_i + h_i) \right)
 \\ & \quad \times \exp i \int dx \Big[ \mathcal{L}_\text{cl}(\bar{g}_i + h_i) 
     +  H_i \frac{\delta \mathcal{L}_\text{cl}(\bar{g}_i + h_i)}{\delta h_i} - \frac{1}{2\alpha} (F_{ij}(\bar{g}) h_j)^2 
     - \frac{1}{\alpha} (F_{ij}(\bar{g}) h_j)(F_{ij}(\bar{g}) H_j) + j_i h_j \Big].
\end{split}
\end{equation}

It is now possible to proceed in two ways. The first option is to integrate over \( H_i \) in Eq.~\eqref{eq27} and then use the resulting \(\delta\)-function to integrate over \( h_i \). This approach is close to what was done with \( Z \) in Eq.~\eqref{eq4} of the preceding section. A second option is to simply derive the Feynman rules for
the propagators and vertices that follow from Eq.~\eqref{eq27} and then consider the possible Feynman diagrams that result.
 
Before examining these two options, we will extend the type of model considered in Eq.~\eqref{eq18}. We will be identifying the field \( g_i(x) \) with the metric \( g_{\mu\nu} \) and the classical Lagrangian \(\mathcal{L}_\text{cl}\) with the Einstein-Hilbert (EH) Lagrangian. By using the LM field \({\lambda}_{\mu\nu}(x)\), we will be able to eliminate the unwanted higher loop divergences that are known to arise when using the EH Lagrangian alone \cite{ref14}. However, we will also include the action for a scalar field interacting with the metric by adding to \(\mathcal{L}_\text{cl}\) a term
\begin{equation} \label{eq28}
    \mathcal{L}_\Phi = \mathcal{L}_\Phi(g_i, \Phi_I).
\end{equation}

We will not employ a LM field to eliminate higher-loop diagrams involving 
the ``matter field'' $\Phi_I$ and that follow from $L_\Phi$ as the divergences 
arising in these diagrams can be absorbed through standard renormalization, 
and if we were to restrict ourselves to one-loop order by use of LM fields, 
unwanted Landau poles will arise in the running coupling \cite{ref14}.

When \( g_i \) undergoes the gauge transformation of Eq.~\eqref{eq21}, then
\begin{equation} \label{eq29}
    \Phi_I \to \Phi_I + R_{Ij}(\Phi) \xi_j.
\end{equation}

If \(\Phi_I\) has a background piece \(\bar{\Phi}_I\) about which there is a quantum fluctuation \(\psi_I\) (\(\Phi_I = \bar{\Phi}_I + \psi_I\)), then \(\Gamma_i\) in Eq.~\eqref{eq27} now depends on \(\bar{g}_i, \bar{\Phi}_I\) (i.e., \(\Gamma = \Gamma(\bar{g}_i, \bar{\Phi}{_I})\)) and the argument of the exponential also has the contributions
\begin{equation} \label{eq30}
\mathcal{L}_\Phi(\bar{g}_i + h_i, \bar{\Phi}{_I} + \psi_I) + k_I \psi_I,
\end{equation}
with
\begin{equation} \label{eq31}
    k_I = -\frac{\delta \Gamma(\bar{g}_i, \bar\Phi_I)}{\delta \bar\Phi_I}.
\end{equation}

After combining Eqs.~\eqref{eq27} and \eqref{eq30}, it is possible to then compute the integral over \(H_i(x)\), leaving us with
\begin{equation}
\begin{aligned}
  \exp\{i \Gamma[\bar{g}_{i}(x), \bar{\Phi}_I(x)]\} &= \int \mathcal{D}h_i \, \mathcal{D}\psi_I \; \det \left(F_{ij}(\bar{g}) R_{jk}(\bar{g} + h_i)\right)
  \det{}^{1/2} \left[\frac{\delta^2 }{\delta h_i(x) \delta h_j(y)}\left(\mathcal{L}_\text{cl}(\bar{g}_i+h_i)-\frac{1}{2\alpha} (F_{ij}(\bar{g}_i) h_j)^2\right) \right]
  \\
    &\qquad \times \exp i\Bigg\{ \int dx \Big[\mathcal{L}_{\text{cl}}(\bar{g}_i + h_i) + \mathcal{L}_{\Phi}(\bar{g}_i + h_i, \Phi_I + \psi_I) 
- \frac{1}{2\alpha} \big(F_{ij}(\bar{g}_i) h_j \big)^2 + j_i h_i + k_I \psi_I \Big]\Bigg\} \\
    &\qquad   \times \delta \left[ \frac{\delta }{\delta h_i} \left(\mathcal{L}_\text{cl}(\bar{g}_i + h_i) - \frac{1}{2\alpha} \big(F_{ij}(\bar{g}_i) h_j\big)^2 \right)\right] .
\end{aligned}
\label{eq32}
\end{equation}
The functional integral over \(h_i\) in Eq.~\eqref{eq32} can be computed using the functional \(\delta\)-function, leading to
\begin{equation} \label{eq33}
    \begin{split}
        \exp \{ i \Gamma(\bar{g}_i, \bar{\Phi}_I(x)) \} ={}& \sum_A \int \mathcal{D}\psi_I {      \det{}^{-1/2} \left[\frac{\delta^2 }{\delta h_i(x) \delta h_j(y)}\left(\mathcal{L}_\text{cl}(\bar{g}_i+h^A_i)-\frac{1}{2\alpha} (F_{ij}(\bar{g}_i) h^A_j)^2\right) \right]
    } 
    \\ & \times
  \det \left( F_{ij}(\bar{g}_i) R_{jk}(\bar{g}_i + h_i^A) \right) 
  \exp i\int dx \bigg[ \mathcal{L}_\text{cl}(\bar{g}_i + h_i^A) + \mathcal{L}_\Phi(\bar{g}_i + h_i^A, \bar{\Phi}_I + \psi_I) 
- \frac{1}{2\alpha} (F_{ij}(\bar{g}) h_j^A)^2
    \\ & 
+ j_i h_i^A + k_I \psi_I \bigg],
\end{split}
\end{equation}
where \(h_i^A\) is a solution to the classical equation of motion
\begin{equation} \label{eq34}
    \frac{\delta}{\delta h_i} \left[ \mathcal{L}_\text{cl}(\bar{g}_i + h_i^A) - \frac{1}{2\alpha} (F_{ij}(\bar{g}) h_j^A)^2 \right] = 0.
\end{equation}

We can identify each of the contributions on the right side of Eq.~\eqref{eq33} with particular sets of Feynman diagrams. The one-loop contributions to the effective action coming from diagrams that exclusively involve the field \(g_i\) come from the term
\(
    {\det}^{-1/2} \left[ \frac{\delta^2}{\delta h_i(x) \delta h_j(y)} \left( \mathcal{L}_\text{cl}(\bar{g}_i + h_i^A) - \frac{1}{2\alpha} (F_{ij}(\bar{g}) h_j^A)^2 \right) \right]
\).
The one-loop contribution of the FP ghost fields is given by
\(
    \det \left( F_{ij}(\bar{g}) R_{jk}(\bar{g}_i + h_i^A) \right)
\).
The tree-level contribution coming from \(\mathcal{L}_\text{cl}(\bar{g} )\) alone all contribute
\(
    \exp i \int dx \left[ \mathcal{L}_\text{cl}(\bar{g}_i + h_i^A) - \frac{1}{2\alpha} (F_{ij}(\bar{g}) h_j^A)^2 \right]
\).
Upon integrating over \(\psi_I\),
\(
    \int \mathcal{D}\psi_I \exp i\int dx \, \mathcal{L}_\Phi(\bar{g}_i + h_i^A, \bar{\Phi}_I + \psi_I)
\)
gives the sum of all diagrams for the field \(\Phi_I\) propagating in a background field \(\bar{g}_i \) and $\bar{\Phi}_I$.

\begin{figure}[ht]
    \centering
    \subfloat[\label{fig:1a} Metric propagator]{\includegraphics[scale=0.7]{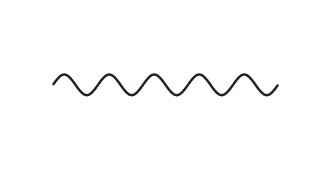}}
    \subfloat[\label{fig:1b} Scalar propagator]{\includegraphics[scale=0.7]{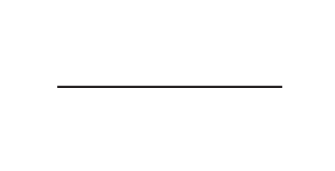}} \\
    \subfloat[\label{fig:1c}  ]{\includegraphics[scale=0.6]{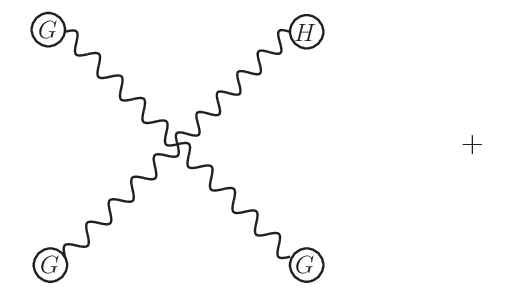}}
    \subfloat[\label{fig:1d}]{\includegraphics[scale=0.6]{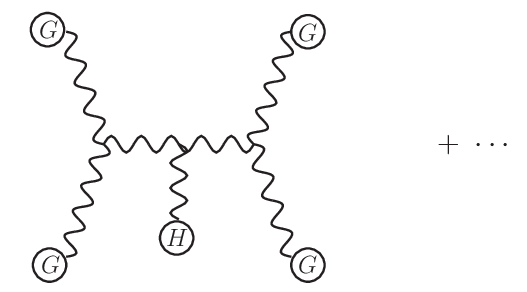}} \\
    \subfloat[\label{fig:1e}]{\includegraphics[scale=0.6]{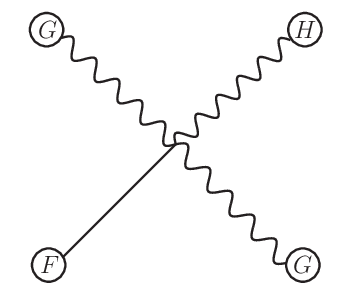}} 
    \subfloat[\label{fig:1f}]{\includegraphics[scale=0.65]{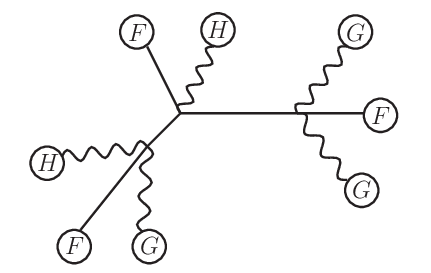}} 
    \caption{Typical tree-level contributions. 
\raisebox{.5pt}{\textcircled{\raisebox{-.5pt} {\scriptsize $G$}}},
\raisebox{.5pt}{\textcircled{\raisebox{-.5pt} {\scriptsize $H$}}},
\raisebox{.5pt}{\textcircled{\raisebox{-.5pt} {\scriptsize $F$}}}
 represents contributions of the fields $ \bar{g}_{\mu \nu} $ and $ h_{\mu \nu}^{A} $, $ \bar{\Phi}_{I} $, respectively.}\label{fig:1}
\end{figure}

\begin{figure}[ht]
    \includegraphics[scale=0.7]{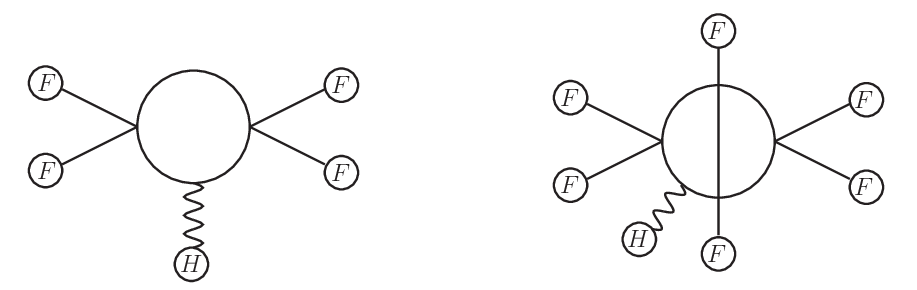}
    \caption{Typical higher loop diagrams from the scalar field.}\label{fig:2}
\end{figure}

Let us consider a diagrammatic approach to these terms appearing in Eq. \eqref{eq33}. If we first make the expansion
\begin{equation} \label{eq:I}
    \begin{split}
        \mathcal{L}_{{\Phi} } ( \bar{g}_{i} + h_{i}^{A} , \bar{\Phi}_{I} + \psi_{I} ) ={}&\mathcal{L}_{\Phi} ( \bar{g}_{i} + h_{i}^{A} , \bar{\Phi}_{I} ) + \int \mathop{d x} \left ( \frac{\partial \mathcal{L}_{\Phi} }{\partial \bar{\Phi}_{I} (x)} \mathcal{L}_{\Phi} ( \bar{g}_{I} + h_{I}^{A} , \bar{\Phi}_{I} ) \right ) \psi_{I} (x) \\ & + \frac{1}{2!} \iint \mathop{d x} \mathop{d y} \left ( \frac{\partial^{2} \mathcal{L}_{\Phi}}{\partial \bar{\Phi}_{I} (x) \partial \bar{\Phi}_{J}(y)}  \mathcal{L}_{\Phi} ( \bar{g}_{I} + h_{i}^{A} , \bar{\Phi}_{I})\right ) \psi_{I} (x) \psi_{J} (y) + \cdots \end{split}
\end{equation}
then the term 
\begin{equation}\label{eq:II}
    \exp i \int \mathop{d x} \left [ \mathcal{L}_\text{cl} ( \bar{g}_{i} + h_{i}^{A} ) - \frac{1}{2 \alpha} ( F_{ij} ( \bar{g ) h_{j}^{A}} )^{2} + \mathcal{L}_{\Phi} ( \bar{g}_{i} + h_{i}^{A} , \bar{\Phi}_{I} ) \right ]
\end{equation}
represents tree-level diagrams \cite{A, B}. In particular, Fig.~\ref{fig:1a} is the propagator following from $ \mathcal{L}_\text{cl} (  \bar{g}_i + h_i^A ) - ( F_{ij} h_{j}^{A} )^{2}/2 \alpha  $ when considering terms quadratic in $ h_{i}^{A} $ when $ \bar{g}_{i} =0$; interactions in this expression lead to tree-level diagrams with 
\raisebox{.5pt}{\textcircled{\raisebox{-.5pt} {\scriptsize $G$}}}
and 
\raisebox{.5pt}{\textcircled{\raisebox{-.5pt} {\scriptsize $H$}}}
representing contributions proportional to $ \bar{g}_{\mu \nu} $ and $ h_{\mu \nu}^{A} $. Figs.~\ref{fig:1c} and~\ref{fig:1d} are typical of such diagrams. The propagator in Fig.~\ref{fig:1b} follows from the term in $ \mathcal{L}_{\Phi} (  \bar{g}_i + h_i^A, \bar{\Phi}_{I} )$ that is bilinear in $ \bar{\Phi}_{I} $ when $ \bar{\Phi}_{I} =  \bar{g}_i + h_i^A = 0 $. Terms in Eq.~\eqref{eq:I} that are at least cubic in $ \psi_{I} $ lead to diagrams such as the tree-level diagrams of Fig.~\ref{fig:1f}, where 
\raisebox{.5pt}{\textcircled{\raisebox{-.5pt} {\scriptsize $F$}}} represents a term proportional to $ \bar{\Phi}_{I} $. 
Fig.~\ref{fig:1e} comes from a term in $ \mathcal{L}_{\Phi} ( \bar{g}_{i} + h^{A}_{i} , \bar{\Phi}_{I} )$ that is quadratic in $ \bar{g}_{i} $ and linear in $ h^{A}_{i} $ and $ \bar{\Phi}_{I} $.  
It is also possible to construct higher loop diagrams in the metric field only contributing as an external field and interior propagators being that of the field $ \Phi_{I} $. Such diagrams appear in Fig.~\ref{fig:2}. 

\begin{figure}[ht]
    \includegraphics[scale=0.8]{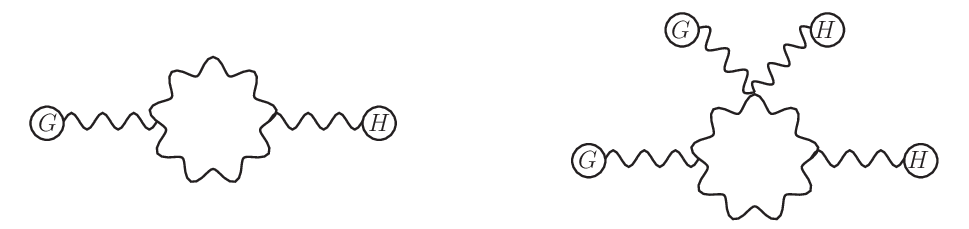}
    \caption{Typical one-loop diagrams with a propagating metric.}\label{fig:3}
\end{figure}

\begin{figure}[ht]
    \includegraphics[scale=0.8]{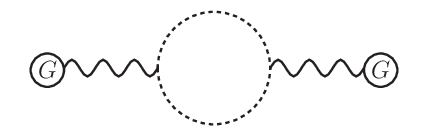}
    \caption{Typical one-loop diagrams from the FP ghost sector.}\label{fig:4}
\end{figure}

Finally, in Eq.~\eqref{eq33}, 
   \( \det{}^{-1/2} \left [ \frac{\delta^{2}}{\delta h_{i} (x) \delta h_{j} (y)} \left ( \mathcal{L}_{\text{cl}} (  \bar{g}_i + h_i^A ) - \frac{1}{2 \alpha} \left ( F_{ij} ( \bar{g}_{i} ) h_{j}^{A}\right )\right ) \right ]\)
   leads to one-loop diagrams with the metric propagating on internal lines and $ \bar{g}_{i} $ and $ h_{i}^{A} $ appearing as external fields. Similarly, $ \det \left( F_{ij} ( \bar{g}_{i} ) R_{jk} (  \bar{g}_i + h_i^A ) \right) $ gives one-loop ghost field contributions with $ \bar{g}_{i} $ and $ h_{i}^{A} $ being external fields. These two types of diagrams are illustrated in Figs.~\ref{fig:3} and Fig.~\ref{fig:4} respectively. The only diagram with contributions beyond one-loop only have internal propagators resulting from the scalar field. Of particular interest is the way in which the propagation of the scalar $ \Phi_{i} $ is affected by $h_{i}^{A} $ which is a solution of the equation of motion of the gravitational field not coupled to $ \Phi_{I} $ itself (ie. Eq.~\eqref{eq34}).

Again, we set the background field associated with the LM equal to zero, as we are only interested in GF with external fields $ g_{i} $. Terms such as $ m_{ijkl} \bar{g}_{i} h_{j} h_{k} h_{l} $, then vanish and so the stability of the action due to terms of order $ h_{i} h_{j} h_{l} $ not lead to an unstable vacuum.

It is now possible to see how these results follow from the Feynman diagrams coming from the path integral arising from
the expression for \(\Gamma\) that follows from Eqs.~\eqref{eq27} and \eqref{eq30}. We will see though that the contribution of $h_i^A$ to the background $ \bar{g}_i$ does not arise from Feynman diagrams alone. (Perturbation theory using only Feynman diagrams is a limited way of examining the effective action. It does not, for example, reveal the effect of instantons on the effective action for Yang-Mills theory.) If we expand \(\mathcal{L}_\text{cl}\) and \(\mathcal{L}_\Phi\) as follows
\begin{subequations}\label{eq35}
\begin{equation} \label{eq35a}
    \mathcal{L}_\text{cl}(\bar{g}_i + h_i) = \mathcal{L}_\text{cl}(\bar{g}_i) + \mathcal{L}_{\text{cl}, i}(\bar{g}_i) h_i + \frac{1}{2!} \mathcal{L}_{\text{cl},ij}(\bar{g}_i) h_i h_j + \cdots,
\end{equation}
\begin{equation} \label{eq35b}
    \mathcal{L}_\Phi(\bar{g}_i + h_i, \bar{\Phi}_I + \psi_I) = \mathcal{L}_\Phi(\bar{g}_i, \bar{\Phi}_I) + \frac{\partial \mathcal{L}_\Phi(\bar{g}_i, \bar{\Phi}_I)}{\partial g_i} h_i + \frac{\partial \mathcal{L}_\Phi(\bar{g}_i, \bar{\Phi}_I)}{\partial \bar{\Phi}_I} \psi_I + \cdots ,
\end{equation}
\end{subequations}
then the terms in the effective Lagrangian that are bilinear in the quantum fields \(\psi_I, h_i,\) and \(H_i\) are of the form
\begin{equation}\label{eq36}
    \begin{pmatrix}
        \psi_I & h_i & H_i
    \end{pmatrix}
    \begin{pmatrix}
        c_{IJ} & b_{Ij} & 0 \\
        b_{iJ} & a_{ij} & a_{ij} \\
        0 & a_{ij} & 0
    \end{pmatrix}
    \begin{pmatrix}
        \psi_J \\ h_j \\ H_j
    \end{pmatrix}.
\end{equation}

There are no terms bilinear in \(H-H\) or \(\psi-H\). By inverting the matrix in Eq.~\eqref{eq36}, we can determine the propagators for these fields,
\begin{equation} \label{eq37}
\begin{pmatrix}
\mathbf{c} & \mathbf{b} & 0 \\[6pt]
\mathbf{b} & \mathbf{a} & \mathbf{a} \\[6pt]
0 & \mathbf{a} & 0
\end{pmatrix}
^{-1}
= \begin{pmatrix}
\mathbf{c}^{-1} & 0 & -\mathbf{c}^{-1} \mathbf{b} \mathbf{a}^{-1} \\[6pt]
0 & 0 & \mathbf{a}^{-1} \\[6pt]
-\mathbf{a}^{-1} \mathbf{b} \mathbf{c}^{-1} & \mathbf{a}^{-1} & -\mathbf{a}^{-1} + \mathbf{a}^{-1} \mathbf{b} \mathbf{c}^{-1} \mathbf{b} \mathbf{a}^{-1}
\end{pmatrix}.
\end{equation}
From Eq.~\eqref{eq37}, we see that the propagators associated with \(\langle h_i h_j \rangle\) and \(\langle h_i \psi_J \rangle\) (or \(\langle \psi_I h_j \rangle\)) vanish, while the ones associated with \(\langle \psi_I(x) \psi_J(y) \rangle\) and \(\langle H_i(x) h_j(y) \rangle\) (or \(\langle h_i(x) H_j(y) \rangle\)) are \(c_{IJ}^{-1}\) and \(a_{ij}^{-1}\), respectively. But now we note that from Eqs.~\eqref{eq27} and \eqref{eq30}, that in any vertex involving the field \(H_i(x)\), that field enters at most linearly.

From these observations, it follows that the only Feynman diagrams that can arise from Eqs.~\eqref{eq27} and \eqref{eq30} are tree-level diagrams that follow from \(\mathcal{L}_\text{cl}\) alone, one-loop diagrams that follow from \(\mathcal{L}_\text{cl}\) alone, and diagrams to any order in the loop expansion that involve propagation of the field \(\psi_I\) in the presence of the background fields \(\bar{g}_i\) and \(\bar{\Phi}_I\) \cite{ref14, ref18}.

This result coincides with what we had obtained in Eq.~\eqref{eq33}, provided that in Eq.~\eqref{eq33} the sum over \(A\) were to include only the contribution \(h_i^A = 0\). In this case, Eq.~\eqref{eq34} reduces to the usual classical equations of motion for the gravitational field. Other solutions to Eq.~\eqref{eq34} do not arise from the Feynman rules in perturbation theory. 
As noted above, these extra solutions to the classical field equations may possibly be associated with non-perturbative aspects of gravity. Such non-perturbative solutions of Eq.~\eqref{eq34} may mimic ``dark matter'' effects \cite{newref29}, as they provide a background to the ``matter field'' $\Phi_I$.

 \section{Divergences with quantized gravitational and scalar fields}\label{sec:divergences}
 
In this section, we show how the approach outlined in Section III can be applied to a scalar field in the presence of a metric \(g_{\mu\nu}\) with the EH Lagrangian.

We consider the 't Hooft and Veltman model (see Refs.~\cite{ref1, ref2}), and we will use the notation used there. The Lagrangian for the metric \(g_{\mu\nu}\) is
\begin{equation} \label{eq38}
    \mathcal{L}_\text{EH} = -\sqrt{g} \, R,
\end{equation}
and for the scalar \(\Phi\) is
\begin{equation} \label{eq39}
    \mathcal{L}_\Phi = -\sqrt{g} \, g^{\mu\nu} \left( \partial_\mu \Phi \right) \left( \partial_\nu \Phi \right).
\end{equation}

The background fields \(\bar{g}_{\mu\nu}\) and \(\bar{\lambda}_{\mu\nu}\) are associated with \(g_{\mu\nu}\) and the LM field \(\lambda_{\mu\nu}\), respectively, with
\begin{equation} \label{eq40}
    g_{\mu\nu} = \bar{g}_{\mu\nu} + h_{\mu\nu}, \quad \lambda_{\mu\nu} = \bar{\lambda}_{\mu\nu} + H_{\mu\nu}.
\end{equation}

According to Eq.~(4.24) of ref.~\cite{ref1}, the term in Eq.~\eqref{eq38} needed for \(\dfrac{\delta^2 \mathcal{L}_\text{cl}(\bar{g} + h)}{\delta h_i(x) \delta h_j(y)}\) is given by
\begin{equation}\label{eq41}
\begin{split}
    \mathcal{L}_\text{EH} (\phi = \bar{\phi} = 0, {g}_{\mu\nu} \to \bar{g}_{\mu\nu} + h_{\mu\nu}^A)
    = \sqrt{\bar{g}} \bigg[&
    - \left(\frac{1}{8}  (h^\alpha_{\ \alpha})^2  - \frac{1}{4} h^\alpha_{\ \beta} h^\beta_{\ \alpha} \right)\bar{R}
- \frac{1}{2} h^{\alpha}_{\ \alpha} \left( h_{\ \beta \ \ \mu}^{\beta\  ; \mu}- h_{\ \mu\ \  \beta}^{\beta\  ; \mu }
-\bar{R}^\beta_{\  \nu} h^\nu_{\  \beta}\right)
\\ & 
 +\frac{1}{ 2}\bar{D}_\alpha\left(
 h^\nu_{\  \mu} h^{\mu  \ ; \alpha}_{\  \nu}
 \right)
 -\frac{1}{ 2}\bar{D}_\beta\left[
 h^\beta_{\ \ \nu}\left( 
2 h^{\nu \alpha}_{\ \ ; \alpha} - h^{\alpha \ \ \nu}_{\ ; \alpha}\right)
 \right]
 \\ & 
    -\frac{1}{4} 
    \left( h^\nu_{\ \beta  ; \alpha} + h^\nu_{\ \alpha ; \beta} - h_{\beta \alpha}^{\ \ \ ; \nu} \right)
    \left( h^{\beta\  ; \alpha}_{\ \nu} + h^{\beta \alpha}_{\ \ ; \nu} - h^{\alpha\  ; \beta}_{\ \nu} \right)
    \\ & 
    +\frac{1}{4}  
    \left( 2 h^{\nu \alpha}_{\ \  ; \alpha} - h^{\alpha \  ; \nu}_{\ \alpha} \right) h^\beta_{\ \beta ; \nu} 
    + \frac{1}{2} h^{\nu \alpha} h^\beta_{\ \ \beta ; \nu \alpha}
  \\ & -\frac{1}{2} h^\nu_{\ \alpha} \bar{D}_\beta 
\left( h^{\beta \  ; \alpha}_{\ \ \nu} + h^{\beta  \alpha}_{\ \ ; \nu} - h^{\alpha \   ; \beta}_{\ \nu} \right)
- h^\nu_{\ \beta} h^\beta_{\ \alpha} \bar{R}^\alpha_{\ \nu}
  \bigg].
  \end{split}
\end{equation}
In Eq.~\eqref{eq41}, both the semicolon ${}_{; \alpha}$ and \(\bar{D}_\alpha\) denote covariant differentiation using the background field \(\bar{g}_{\mu\nu} + h_{\mu\nu}^A\),
\begin{equation} \label{eq42}
    V_{\beta ; \alpha} \equiv \bar{D}_\alpha V_\beta = \partial_\alpha V_\beta - \bar{\Gamma}^\nu_{\alpha\beta} V_\nu,
\end{equation}
and $ \bar{R} = \bar{R}_{\mu \nu} \bar{g}^{\mu \nu}  $, $ \bar{R}_{\mu \nu} = \bar{R}^{\alpha }_{\mu \alpha \nu} $ with 
\begin{equation} \label{eq43}
    \bar{R}^\mu_{\nu\alpha\beta} = \partial_\alpha \bar{\Gamma}^\mu_{\beta\nu} - \partial_\beta \bar{\Gamma}^\mu_{\alpha\nu} + \bar{\Gamma}^\mu_{\alpha\lambda} \bar{\Gamma}^\lambda_{\beta\nu} - \bar{\Gamma}^\mu_{\beta\lambda} \bar{\Gamma}^\lambda_{\alpha\nu}.
\end{equation}

By Eq.~(4.22) of ref.~\cite{ref17}, the term in Eq.~\eqref{eq34} needed for
\(\delta \mathcal{L}_\text{cl} (\bar{g}_{i} + h_{i}^A)/\delta h_i \) is
\begin{equation} \label{eq44}
 \underline{{\cal L}}(\phi=\bar\phi=0,{g}_{\mu\nu}\to\bar{g}_{\mu\nu}+h^A_{\mu\nu} )   
 = \sqrt{\bar{g}} \left( -\frac{1}{2} h_{\ \alpha}^\alpha \bar{R} + h^\beta_{\ \alpha}  \bar{R}^\alpha_{\ \beta} -h^{\beta}_{\ \beta}{}_{; \alpha}^{\ \  \alpha}+ h^{\alpha\beta}_{\ \ \ ; \alpha\beta}\right).
\end{equation}
In  Ref.~\cite{ref1}, (see Eq. 4.29) a gauge-fixing condition that respects transformations corresponding to diffeomorphism of the background fields $\bar{g}_{\mu\nu}$ and $\Phi$ is employed.
This condition depends explicitly on both fields. 
(A discussion of gauge conditions that respect this invariance is in ref.~\cite{ref24}.) Instead, the gauge-fixing
\begin{equation} \label{eq45}
    \mathcal{L}_\text{gf} = -\frac{1}{2} \sqrt{\bar{g}} \left( h_{\mu \  ; \nu}^{\ \nu} - \frac{1}{2} h^{\nu}_{\ \nu ; \mu} \right) \left( h^{\mu\lambda}_{\ \ \ ; \lambda} - \frac{1}{2} h^{\lambda\ \ ; \mu}_{\ \lambda} \right)
\end{equation}
is used. The independence of physical results on the gauge-fixing condition is discussed in \cite{ref25}. The FP Lagrangian for the fermionic ghost fields \(\eta_\mu\) and \(\eta_\mu^\star\) associated with Eq.~\eqref{eq45} is (Eq.~(5.19) of ref.~\cite{ref1})
\begin{equation} \label{eq46}
    \mathcal{L}_\text{ghost} = \sqrt{\bar{g}} \, \eta^\star_\mu \, \left( \eta^{\mu ;\alpha}_{\ \ \ \alpha}  - \bar{R}^{\mu\alpha} \eta_\alpha \right) .
\end{equation}
In Eqs.~\eqref{eq41}, \eqref{eq44} and \eqref{eq46}, $g$ denotes $ \det (\bar{g}_{\mu\nu} +  h_{\mu\nu}^A)$ and $\bar{R}$ denotes the Ricci scalar evaluated at $\bar{g}_{\mu\nu} + h_{\mu \nu}^{A} $.

Having seen how a LM field can be used to eliminate diagrams involving the graviton field in diagrams beyond one-loop order, we can examine the divergences arising in the generating functional of Eq.~\eqref{eq33}.


The divergences that arise in models involving scalars, vector, and spinor fields in flat spacetime can, under many circumstances, be absorbed by renormalizing the fields themselves, their masses, or the couplings characterizing their interactions. This is a feature of the so-called ``Standard Model'' of particle interaction. (See, for example, ref.~\cite{ref26}.)

Divergences also arise when computing radiative effects in curved spacetime. (Three useful references are \cite{ref27, ref28, ref29}.) There are now also divergences associated with radiative effects resulting from the propagation of the metric field. If we follow the argument following Eq.~\eqref{eq27} and apply it to that portion of Eq.~\eqref{eq32} that does not involve the matter field \(\Phi_I = \bar{\Phi}_I + \psi_I\), then we see that the only diagrams involving the propagation of the metric (graviton) field are the one-loop diagrams that follow from Eqs.~\eqref{eq38}, \eqref{eq45}, and \eqref{eq46}. 

The divergences arising in these one-loop diagrams were computed in Eq.~(5.24) of ref.~\cite{ref1} to be
\begin{equation} \label{eq47}
    \Delta S_\text{EH div} = \left( \frac{ \sqrt{\bar{g} }}{ \epsilon} \right) \left( \frac{1}{120} \bar{R}^2 + \frac{7}{20} \bar{R}_{\mu\nu} \bar{R}^{\mu\nu} \right),
\end{equation}
where the metric is evaluated at the sum of \(\bar{g}_{\mu\nu}\) (the full background metric) and \(h_{\mu\nu}^A\), which is a solution to the equation of motion for the metric in the presence of the background field \(g_{\mu\nu}\), using the EH alone to determine these equations of motion (Eq.~\eqref{eq34}). (In Eq.~\eqref{eq47}, \(\epsilon = 8\pi^2 (n - 4)\), where \(n\) is the number of dimensions.)

There are also divergences resulting from the remaining contributions to the generating functional \(\Gamma\) in Eq.~\eqref{eq32}.
These remaining portions are given by the path integral
\begin{equation} \label{eq48}
    \int \mathcal{D}\psi_I \, \exp \left\{ i\int dx \, \mathcal{L}_\Phi (\bar{g}_{i} + h_{i}^A,  \bar\Phi_I + \psi_I) \right\}.
\end{equation}

This represents all Feynman diagrams for a scalar field \(\Phi_I\) with background \(\bar{\Phi}_I\), propagating in a background metric \(\bar{g}_{\mu\nu} + h_{\mu\nu}^A\), where \(\bar{g}_{\mu\nu}\) is a general background metric and \(h_{\mu\nu}^A\) is a solution to the equation \eqref{eq34}. This equation of motion involves solely the EH Lagrangian; it does not have any contribution from the scalar field.

A discussion of the divergences arising from the propagation of a scalar field in the presence of a background metric (in this case, \(\bar{g}_{\mu\nu} + h_{\mu\nu}^A\)) appears in refs.~\cite{ref28, ref29}. The action there consists not only of the free part of Eq.~\eqref{eq34} but, in addition, a mass term and self-interaction term. We also include a cosmological constant term. In total, we have
\begin{eqnarray} \label{eq49}
    S_\Phi &=& \int dx \, \mathcal{L}_\Phi (\bar{g}_{\mu\nu} + h_{\mu\nu}^A, \bar{\Phi} + \psi)
 \nonumber \\
    &=& \int dx \, \sqrt{g} \left( 
    -\frac{1}{2} g^{\mu\nu} \partial_\mu \phi \partial_\nu \phi
    + \frac{1}{2} \left( \xi \, R - m^2 \right) \phi^2
    - \frac{\lambda}{4!} \phi^4
    + \Lambda \right).
\end{eqnarray}

Divergences arise both from diagrams in which the only external field is the background metric (the ``vacuum diagrams'') and also from diagrams in which there is a background scalar field, as well as possibly contributions from the external metric.
This external metric is denoted by $ \bar{g}_{\mu \nu} $. The ``vacuum diagrams'' do not contribute when the background metric is flat.

The ``vacuum diagrams'' result in divergences when \(\epsilon = 0\) that are proportional to (Eq.~(3.45) of Ref.~\cite{ref28})
\begin{equation} \label{eq50}
    \sqrt{g},
\end{equation}
or are proportional to
\begin{subequations}\label{eq51a-e}
\begin{equation} 
    \sqrt{g}\, R, 
    \quad 
 \sqrt{g} R^2,
 \quad 
 \sqrt{g} R_{\mu\nu} R^{\mu\nu},
 \quad 
 \sqrt{g} R_{\mu\nu\alpha\beta} R^{\mu\nu\alpha\beta}
 \quad \text{or }
 \quad 
 \sqrt{g} \, \, \Box R,
\end{equation}
\end{subequations}
where $ \Box \equiv \partial^{\mu} \partial_{\mu} $.

The divergences at \(\epsilon = 0\) that arise from Feynman diagrams involving external scalar fields are proportional to
\begin{subequations}\label{eq52}
\begin{equation}\label{eq52a} 
  \sqrt{g} g^{\mu\nu} \partial_\mu \phi \partial_\nu \phi,
  \quad 
  \sqrt{g} \phi^2 ,
  \quad 
  \sqrt{g} \phi^4 ;
\end{equation}
\begin{equation}\label{eq52b} 
  \sqrt{g} R \phi^2.
\end{equation}

\end{subequations}

It is possible to systematically remove all the divergences in Eqs.~\eqref{eq47}, \eqref{eq50}–\eqref{eq52}. First, divergences involving terms of the form of Eq.~\eqref{eq52a} can be absorbed by the usual field strength, mass, and coupling constant renormalizations, respectively. Meanwhile, divergences arising from terms proportional to \(\sqrt{g}\) (Eq.~\eqref{eq50}) can be absorbed by the cosmological constant \(\Lambda\). This holds to all orders in the loop expansion of contributions to the effective action that follow from Eq.~\eqref{eq48}.

The remaining divergences (those of Eqs.~\eqref{eq47}, \eqref{eq51a-e}, \eqref{eq52b}) are all proportional to \(R_{\mu\nu}\), since \(R= g^{\mu\nu}R_{\mu\nu}\), and in four dimensions there is the topological identity
\begin{equation} \label{eq53}
    R_{\mu\nu\alpha\beta} R^{\mu\nu\alpha\beta} = -R^2 + 4 R_{\mu\nu} R^{\mu\nu}.
\end{equation}

With the classical Lagrangian \(\mathcal{L}_\text{cl}\) being given by \(\mathcal{L}_\text{EH}\) in Eq.~\eqref{eq38}, the term
\(
    \lambda_i \frac{\delta \mathcal{L}_\text{cl}}{\delta g_i}
\)
in Eq.~\eqref{eq18} becomes
\begin{eqnarray} \label{eq54}
  \lambda_{\mu\nu}   \frac{\delta}{\delta g_{\mu\nu}} \left( -\sqrt{g} R \right)
  &=& \lambda_{\mu\nu} \frac{\sqrt{g}}{2} \left( R^{\mu\nu} - \frac{1}{2} R g^{\mu\nu} \right)
\nonumber \\
  &=& \left( \lambda^{\mu\nu} - \frac{1}{2} g^{\mu\nu} g^{\alpha\beta} \lambda_{\alpha\beta}  \right) \frac{\sqrt{g}}{2} R_{\mu\nu} \equiv \frac{ \sqrt{g}}{2} \Lambda^{ \mu \nu} R_{\mu \nu},
\end{eqnarray}
where we have set 
\begin{equation} \label{eq55}
    {\Lambda}^{\mu\nu} = \lambda^{\mu\nu} - \frac{1}{2} g^{\mu\nu} \lambda, \quad 
\lambda \equiv  \, {g}^{\alpha\beta} \lambda_{\alpha\beta} .
\end{equation}
Thus, after renormalizing the $ \Lambda^{\mu \nu} $ field as
\begin{equation}\label{eq:4.19}
    \Lambda^{\mu \nu}  = \Lambda^{\mu \nu}_{\text{ren}} + c_{1} g^{\mu \nu} R + c_{2} R^{\mu \nu}
\end{equation}
one can see from Eq.~\eqref{eq54} that the divergences in the effective action \eqref{eq47} which involves external metric fields can be absorbed in the coefficients $c_1$ and $ c_2$.

There is no need to introduce extra terms into the classical action (as was done in refs.~\cite{ref12, ref13}) to eliminate divergences that arise at higher loop orders when considering the Einstein-Hilbert action of Eq.~\eqref{eq38} alone \cite{ref6, ref7}. Such terms involve having propagators that are quartic in the momentum transfer, which compromises unitarity. When a Lagrange multiplier (LM) field is used to eliminate higher-loop divergences, BRST invariance is intact, and so the resulting theory is unitary \cite{ref18}. This we further discuss in an appendix below.

\section{Discussion}

In this work, we considered, for simplicity, the renormalization of a theory involving a single scalar field interacting with a gravitational field. Using the LM approach, we evaluated the effective action in closed form. We have shown that extra solutions to the Einstein classical equation of motion, which contribute to the effective action, arise. We have established that, in the presence of a LM field, the radiative corrections arising from internal metric fields are restricted to one-loop order.
The only diagrams beyond one loop order have solely matter fields propagators.
The renormalization group equation, which reflects the possibility of choosing generic renormalization scales, could lead to interesting insights into the behavior of the model in the ultraviolet regime. 
It should be possible to apply the same approach, which maintains unitarity, if there were also multiple scalar or non-Abelian vector gauge fields coupled to the gravitational field. 

If spin-$\frac{1}{2}$ matter fields were to couple to the metric field, then one must consider the Einstein-Cartan  action in place of the EH action of Eq.~(38) \cite{ref30}. The Einstein-Cartan action possesses two distinct local gauge invariances and consequently, its quantization is more involved than quantizing the EH action \cite{ref31,ref32}. The problem of using a LM field to eliminate higher loop diagrams in a model involving spin-$\frac{1}{2}$ fields coupled to the metric is currently being considered. If the couplings can be shown to be consistent with unitarity and renormalizability, it may be possible to consistently extend the Standard Model to incorporate gravity.


\begin{acknowledgments}
D.\ G.\ C\@. M\@. thanks Ann Aksoy for enlightening conversations and Farrukh Chishtie for useful correspondence. 
F.\ T.\ B\@., J.\ F\@. thank CNPq (Brazil) for financial support. 
S.\ M.-F\@. thank CNPq (Brazil) for partial financial support.
This study was financed, in part, by the São Paulo Research Foundation (FAPESP), Brasil. Process Number: 2024/19216-8. 
\end{acknowledgments}

\appendix
\section{Unitarity with LM fields}

In order to illustrate the unitarity with the presence of LM fields, we consider a simple model described by the Lagrangian 
with a scalar field $ A$ and a LM field $B$ :
\begin{equation}\label{eq:A1}
\mathcal{L} = - \frac{1}{2} ( A \Box A + m^2 A^2) - \frac{g}{4!} A^4 - ( B \Box A + m^2 AB)- \frac{g}{3!} A^3 B
.
\end{equation}
We note that there appears in the quadratic part of the Lagrangian a mixing between the field $ A $ and $ B $. In order to diagonalize this part, we introduce a new field $C$ defined by $ C = A + B $ so that \eqref{eq:A1} may be written in the form 
\begin{equation}\label{eq:A2}
\mathcal{L} =- \frac{1}{2} ( C \Box C + m^2 C^2) + \frac{1}{2} ( B \Box B + m^2 B^2)  - \frac{g}{3!} B (C-B)^3 - \frac{g}{4!} (C-B)^4
,
\end{equation}
where the two-point functions involving the $A$ and $C$ fields are decoupled, though now the self coupling of $B$ and $C$ are more complicated. 

Following Ref.~\cite{ref18}, we split these fields into a background and quantum part: 
\begin{equation}\label{eq:A3}
C = \bar{C} + \mathcal{C}, \quad B = \bar{B} + \mathcal{B}
.
\end{equation}
Identifying the solutions of the classical equations of motion with the background fields, we obtain the equations
\begin{subequations}
\begin{align}\label{eq:A4}
( \Box  + m^2 ) \bar{C} &= - \frac{g}{6} (\bar{C}-\bar{B})^3
- \frac{g}{2} \bar{B} (\bar{C}-\bar{B})^2
,
\\ \label{eq:A5}
( \Box  + m^2 ) \bar{B} &= 
- \frac{g}{2} \bar{B} (\bar{C}-\bar{B})^2
.
\end{align}
\end{subequations}
In order to preserve the original equation of motion, one must have $  \bar{B}  = 0$, in which case \eqref{eq:A4} becomes: 
\begin{equation}\label{eq:A6}
( \Box  + m^2 ) \bar{C} = - \frac{g}{6} \bar{C}^3
.
\end{equation}
Thus, using \eqref{eq:A3}, we see that the LM field is a purely quantum field (this is discussed in the case of a gauge theory in Ref. \cite{ref33}).

\begin{figure}[h]
 \centering
    \subfloat[\label{fig:0c} Scalar quartic vertex.]{\includegraphics[scale=0.4]{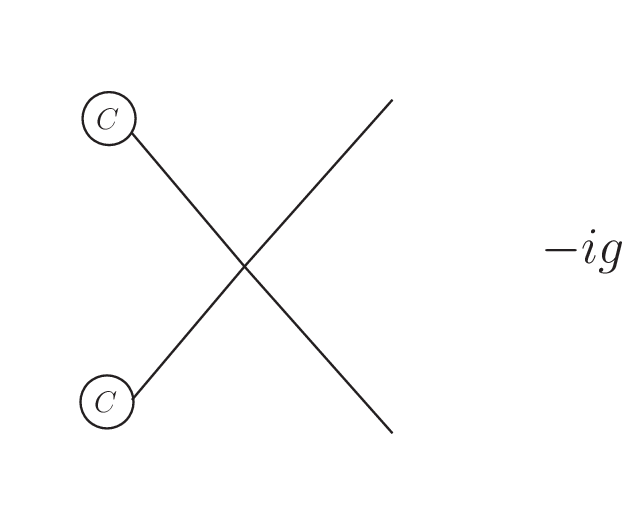}}
    \subfloat[\label{fig:0d} $ \mathcal{B}^2 $ quartic vertex.]{\includegraphics[scale=0.4]{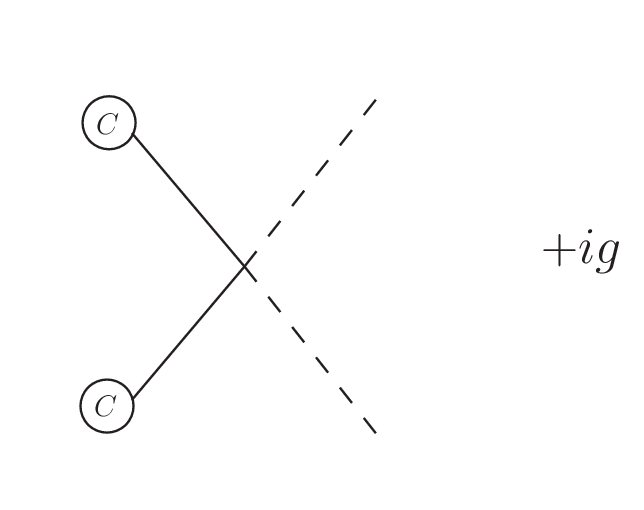}}
    \subfloat[\label{fig:0e} $ \mathcal{B}^{3} $ quartic vertex.]{\includegraphics[scale=0.43]{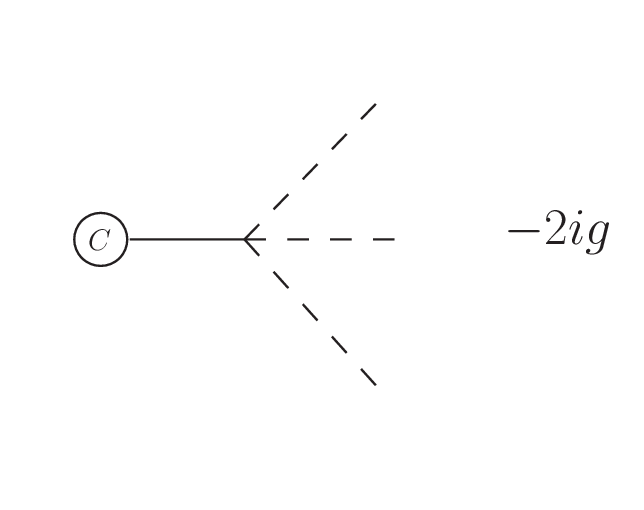}}
    \caption{Feynman rules of the vertices coming from the Lagrangian \eqref{eq:A2} using the background field method. Solid lines denote the quantum field $ \mathcal{C} $, while dashed lines represent the quantum field $ \mathcal{B} $. 
\raisebox{.5pt}{\textcircled{\raisebox{-.5pt} {\scriptsize $C$}}}
     represents contributions of the background field $\bar{C}$.}\label{fig:A0}
\end{figure}

Moreover, in \eqref{eq:A2}, the part quadratic in the quantum fields $ \mathcal{C} $ and $ \mathcal{B} $ leads to the propagators: 
\begin{equation}\label{eq:A7}
D_{ \mathcal{C}\mathcal{C}} (k) = +\frac{i}{k^2 - m^2 + i\epsilon}  \quad
D_{ \mathcal{B}\mathcal{B}} (k) = -\frac{i}{k^2 - m^2 + i \epsilon};
\end{equation}
while the part involving the interaction of the fields is given by 
\begin{equation}\label{eq:A8}
\mathcal{L}_{\text{int}} = - \frac{g}{4!} (\bar{C} + \mathcal{C})^4 + \frac{g}{2!2!} (\bar{C} + \mathcal{C})^2 \mathcal{B}^2 - \frac{2 g}{3!} (\bar{C} + \mathcal{C}) \mathcal{B}^3 + \frac{3 g}{4!} \mathcal{B}^4
.
\end{equation}
Using these expressions, one may derive the corresponding Feynman rules (see Fig. \ref{fig:A0}) needed for the evaluation of the Feynman loop diagrams.

\begin{figure}[h]\centering
    \includegraphics[scale=0.5]{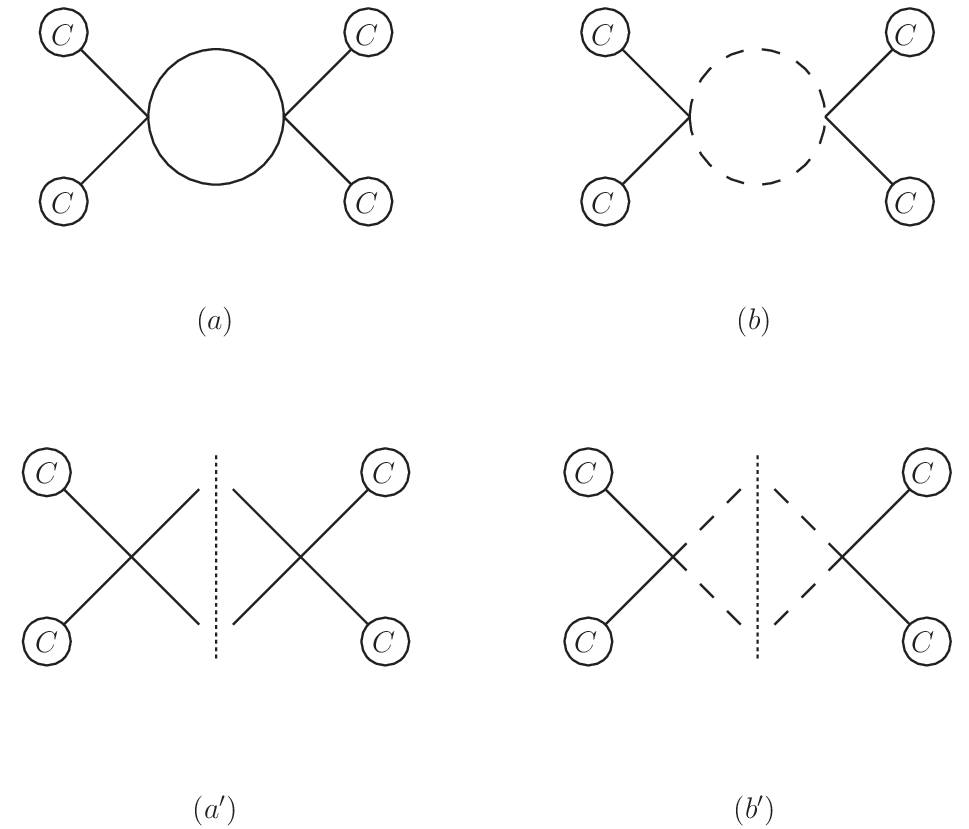}
    \caption{One-particle irreducible one-loop diagrams that contribute to the $ \bar{C} $--$ \bar{C} $  scattering amplitude.}\label{fig:A1}
\end{figure}

Let us consider first the contribution to the $ \bar{C} $--$ \bar{C} $ scattering amplitude involving the background field $\bar{C}$, shown in Figs.\ref{fig:A1}(a, b).
One can verify that the imaginary part of this forward scattering amplitude is proportional to the cross-section corresponding to the process shown in Figs. \ref{fig:A1}(a', b') in accordance to the unitarity requirement.

\begin{figure}[h]\centering
    \includegraphics[scale=0.75]{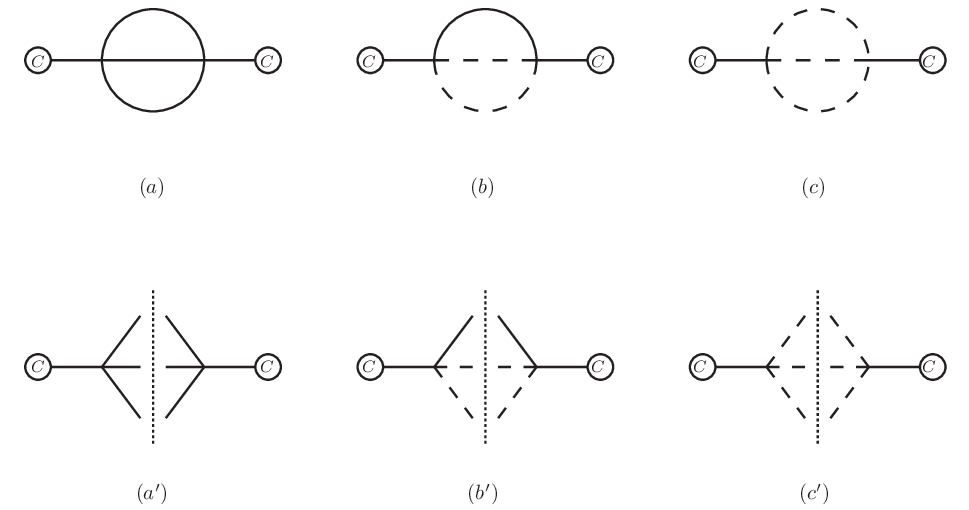}
    \caption{Sunset (two-loop order) diagrams.}\label{fig:A2}
\end{figure}

Next, let us consider the sunset diagrams depicted in Fig. \ref{fig:A2}.
The $2$-loop contributions arising from the diagrams in Figs. \ref{fig:A2}(a, b, c) cancel out, since in the presence of LM there are no net contributions beyond one-loop order. (we refer the reader to the Appendix A of Ref. \cite{ref17} for more detail.) This implies that corresponding imaginary parts should add up to zero. This may be verified by computing the contributions from the diagrams in Figs. \ref{fig:A1}(a', b', c').


The results above confirm the unitarity of the LM theory, which relates cross sections to the imaginary parts of the forward scattering amplitudes, in accordance with the optical theorem. 

\subsection{Extended LM formalism}
When the determinant factor $ \det{ \mathcal{L}_{,ij}}^{1/2}$ is included, which corresponds to the extended LM formalism as outlined in Refs. \cite{ref19, ref20}, unitarity remains intact. Next, we will illustrate it with the scalar model considered above.

Using the background field method, the determinant factor leads to the following Lagrangian
\begin{equation}\label{eq:lagdet}
    \det{}^{1/2} \mathcal{L}^{''} ( \bar{C} ) 
    = \exp \frac{1}{2} \mathop{\rm Tr} \ln{ \left ( 1 + \frac{g}{2} \frac{1}{\Box + m^{2}}\right ) \bar{C}^{2}},
\end{equation}
where we have omitted a field-independent term which may be absorbed in a normalization factor. Expanding Eq.~\eqref{eq:lagdet} in powers of $ g$, we obtain 
\begin{equation}\label{eq:A9}
    \det{}^{1/2} \mathcal{L}^{''} ( \bar{C} ) 
    = 
    \exp \frac{1}{2} \mathop{\rm Tr}  \left[ \frac{g}{2} \frac{1}{ \Box + m^{2}} \bar{C}^{2} - \frac{g^{2} }{8} \left ( \frac{1}{\Box + m^{2} } \bar{C}^{2} \right )^{2} + O(g^{3} )\right].
\end{equation}
The first term in the exponential yields a tadpole contribution which vanishes in dimensional regularization. The second term leads to a contribution which cancels that shown in Figs.~\ref{fig:A1}(b), which arises from the LM fields. Note that in Eq.~\eqref{eq:A9} there are no terms of the form $ g^{2} \bar{C}^{2} $, which correspond to contributions arising from the sunset diagrams shown in Fig.~\ref{fig:A2}. This reflects the fact that such contributions cancel out.


\begin{thebibliography}{99}

\bibitem{ref1} G. 't Hooft and M. Veltman, Ann. Inst. Henri Poincar\'e, \textbf{20}, 69 (1974).

\bibitem{ref2} M. Veltman, in ``Les Houches 1975, Methods in Field Theory,'' (ed. Roger Balian and Jean Zinn-Justin), North Holland Publishing (1976), Amsterdam.

\bibitem{ref3} G. 't Hooft, ``Perturbative Quantum Gravity,'' in ``International School of Subnuclear Physics, 40th Course: From Quarks and Gluons to Quantum Gravity,'' Erice, Sicily, July (2002), pg. 249. \url{https://www.staff.science.uu.nl/~hooft101/lectures/erice02.pdf}.

\bibitem{ref4} S. Deser, H.S. Tsao, and P. van Nieuwenhuizen, Phys. Rev. D \textbf{10}, 3337 (1974).

\bibitem{ref5} S. Deser and P. van Nieuwenhuizen, Phys. Rev. D \textbf{10}, 411 (1974).

\bibitem{ref6} M.H. Goroff and A. Sagnotti, Nucl. Phys. B \textbf{266}, 709 (1980).

\bibitem{ref7} A.E.M. van de Ven, Nucl. Phys. B \textbf{378}, 309 (1992).

\bibitem{ref8} D.Z. Freedman and A. Van Proeyen, ``Supergravity,'' Cambridge University Press, Cambridge, UK (2012).

\bibitem{ref9} P. West, ``Introduction to Strings and Branes,'' Cambridge University Press, Cambridge, UK (2012).

\bibitem{ref10} R. Gambini and J. Pullin, ``A First Course in Loop Quantum Gravity,'' Oxford University Press, Oxford, UK (2011).

\bibitem{ref11} S. Nagy, Ann. Phys. (Amsterdam) \textbf{350}, 310 (2014).

\bibitem{newref17} I.B. Khriplovich, Yad. Phys. \textbf{10} (1969) 409 [Sov. J. Nucl. Phys. \textbf{10} (1970) 23]

\bibitem{ref12} K. Stelle, Phys. Rev. D \textbf{16}, 953 (1977).

\bibitem{ref13} P.D. Mannheim, Found. Phys. \textbf{42}, 388 (2012).

\bibitem{ref14} D.G.C. McKeon and T.N. Sherry, Can. J. Phys. \textbf{70}, 441 (1992).

\bibitem{ref15} F.T. Brandt, J. Frenkel, and D.G.C. McKeon, Can. J. Phys. \textbf{98}, 344 (2020).

\bibitem{ref16} F.T. Brandt, J. Frenkel, D.G.C. McKeon, and G.S.S. Sakoda, Phys. Rev. D \textbf{100}, 125014 (2019).

\bibitem{ref17} F.T. Brandt, J. Frenkel, S. Martins-Filho, and D.G.C. McKeon, Ann. Phys. \textbf{427}, 168426 (2020).

\bibitem{ref18} F.T. Brandt, J. Frenkel, S. Martins-Filho, D.G.C. McKeon, Ann. Phys. \textbf{434}, 168659 (2021).

\bibitem{ref19} F.T. Brandt and S. Martins-Filho, Ann. Phys. \textbf{453}, 169323 (2023).

\bibitem{ref20} F.T. Brandt, S. Martins-Filho, and D.G.C. McKeon, Eur. Phys. J. C \textbf{84}, 399 (2024).

\bibitem{ref36}
M.~Y.~Kalmykov, Class. Quant. Grav. \textbf{12}, 1401-1412 (1995)



\bibitem{ref21} B.S. DeWitt, Phys. Rev. \textbf{162}, 1195 (1967).

\bibitem{ref22} L.F. Abbott, Nucl. Phys. B \textbf{185}, 189 (1981); Acta Phys. Pol. B \textbf{13}, 33 (1982).

\bibitem{ref23} L.D. Faddeev and V.N. Popov, Phys. Lett. B \textbf{25B}, 29 (1967).

\bibitem{A} D.~G.~Boulware and L.~S.~Brown, Phys. Rev. \textbf{172}, 1628-1631 (1968)

\bibitem{B} M.~J.~Duff, Phys. Rev. D \textbf{7}, 2317-2326 (1973)

\bibitem{newref29} L.~Hui, J.~P.~Ostriker, S.~Tremaine and E.~Witten,
Phys. Rev. D \textbf{95}, no.4, 043541 (2017)

\bibitem{ref24} F.T. Brandt, J. Frenkel, and D.G.C. McKeon, Phys. Rev. D \textbf{106}, 065010 (2022).

\bibitem{ref25} G. 't Hooft and M. Veltman, Nucl. Phys. B \textbf{50}, 318 (1972).

\bibitem{ref26} S. Weinberg, ``The Quantum Theory of Fields'' (Vol. I and II), Cambridge University Press, Cambridge, UK (1996).

\bibitem{ref27} N.D. Birrell and P.C.W. Davies, ``Quantum Fields in Curved Space,'' Cambridge University Press, Cambridge, UK (1982).

\bibitem{ref28} J. Buchbinder, S.D. Odintsov, and I.L. Shapiro, ``Effective Action in Quantum Gravity,'' IOP Publishing, Bristol, UK (1992).

\bibitem{ref29} J. Buchbinder and I.L. Shapiro, ``Introduction to Quantum Field Theory with Applications to Quantum Gravity,'' Oxford University Press, Oxford, UK (2021).

\bibitem{ref30} F.W. Hehl, P. van der Heyde, G.D. Kerlick, and J.M. Nester, Rev. Mod. Phys. \textbf{48}, 393 (1976).



\bibitem{ref31} F.T. Brandt, J. Frenkel, S. Martins-Filho, and D.G.C. McKeon, Ann. Phys. \textbf{462}, 169607 (2024).

\bibitem{ref32} F.T. Brandt, J. Frenkel, S. Martins-Filho, and D.G.C. McKeon, Ann. Phys. \textbf{470}, 169801 (2024).

\bibitem{ref33}  S. Martins-Filho, ``\href{https://doi.org/10.11606/T.43.2024.tde-19122024-130256}{Covariant quantization of gauge theories with Lagrange multipliers}'', Doctoral dissertation, Universidade de São Paulo (2024). {\tt arXiv:2504.04666 [hep-th]}


\end{thebibliography}
\end{document}